\documentclass{aa}
\usepackage{graphicx}
\usepackage{subfig}
\usepackage{natbib}
\usepackage{rotating}
\usepackage{multirow}
\usepackage[varg]{txfonts}

\newcommand{\xmm}{{\small \it XMM-Newton}}
\newcommand{\pn}{{\small EPIC-pn}}
\newcommand{\chandra}{{\small \it Chandra}}
\newcommand{\lm}{11~LMi}
\newcommand\nodata{ ~$\cdots$~ }
\newcommand{\iot}{$\iota$~Hor}
\newcommand{\tb}{$\tau$~Boo}
\newcommand{\ac}{$\alpha$~Cen}

\begin{document}

\title{Coronae of stars with supersolar elemental abundances}

\author{Uria Peretz\inst{1} \and Ehud Behar\inst{1} \and Stephen A. Drake\inst{2}}
\institute{Department of Physics, Technion, Haifa 32000, Israel \and USRA/CRESST \& NASA/GSFC, Greenbelt, Maryland 20771, USA}

\abstract{
Coronal elemental abundances are known to deviate from the photospheric values of their parent star, with the degree of deviation depending on the first
 ionization potential (FIP). This study focuses on the coronal composition of stars with supersolar photospheric abundances. We
 present the coronal abundances of six such stars: 11 LMi , $\iota$ Hor, HR 7291, $\tau$ Boo, and $\alpha$ Cen A and B. These stars all have
 high-statistics X-ray spectra, three of which are presented for the first time. The abundances we measured were obtained using
 the line-resolved spectra of the Reflection Grating Spectrometer (RGS) in conjunction with the higher throughput \pn\ camera spectra
 onboard the \xmm\ observatory. A collisionally ionized plasma model with two or three temperature components is found to
 represent the spectra well. All elements are found to be consistently depleted in the coronae compared to their respective photospheres.
 For 11 LMi and $\tau$ Boo no FIP effect is present, while $\iota$ Hor, HR 7291, and $\alpha$ Cen A and B show a clear FIP trend. These
 conclusions hold whether the comparison is made with solar abundances or the individual stellar abundances. Unlike the solar corona, where low-FIP elements are enriched,
 in these stars the FIP effect is consistently due to a depletion of high-FIP elements with respect to actual photospheric abundances. A comparison
 with solar (instead of stellar) abundances yields the same fractionation trend as on the Sun. In both cases, a similar FIP bias is inferred,
 but different fractionation mechanisms need to be invoked.
}


\maketitle

\section{Introduction}
\label{intro}

The active coronae of cool stars can be up to five orders of magnitude more luminous and an order of magnitude hotter than our own solar corona.
Like the Sun, they emit both extreme UV (EUV) and soft X-rays.
X-ray spectroscopy allows for particularly deep insights into the hot thermal content and elemental abundances of the corona. For comprehensive reviews see \citet{Guedel2009} and \citet{testa2010}.
Solar coronal abundances are different from the solar photospheric composition.
The abundance patterns depend on the first ionization potential (FIP) \citep{Meyer1985,Feldman1992}. Elements with FIP$<$10~eV (low FIP) are enriched in the corona relative to elements with FIP$>$10~eV (high FIP). 
Averaging over the entire solar disk, the enrichment factor is about 4 \citep[][ and references therein]{Feldman2002}. 
Currently, there is no generally accepted model that explains the FIP bias \citep[but see][]{Laming04, Telleschi05, Laming12}. Nevertheless, the observed reality of abundance fractionation is undisputed, although the absolute normalization of the solar coronal abundances is still being debated \citep[cf.,][]{schmelz12}.

The FIP effect became even more puzzling when high-resolution X-ray spectra from \xmm\ and \chandra\ revealed that active stellar coronae do not follow the solar FIP pattern. In some cases, the high-FIP elements are even enriched compared to the low-FIP ones, an effect labeled inverse-FIP (IFIP) effect \citep{Brinkman2001}. 
Later studies revealed that FIP and IFIP biases are correlated with coronal activity (e.g., X-ray luminosity) and age: 
Highly active stars show an IFIP effect, while less active coronae have a solar FIP bias \citep{Audard03, Telleschi05}, implying a transition on stellar evolution time-scales  of Gyr.
Intervening intermediate-activity stars seem to exhibit a relatively flat dependence on FIP. 
\citet{Ball05} even suggested a U-shaped dependence, although such subtleties depend on one or two elements at most.  
Some more recent observations of stellar coronae have found that there are deviations from this behavior for some stars, however, and that there could be additional factors that determine coronal abundances, for example, spectral type \citep{Guedel2007, Wood10}.
 In particular, the influence of stellar metallicity on stellar coronal properties is still unclear, since most of the stars with well-studied X-ray spectra have photospheric abundances similar (or assumed to be) to the solar photospheric values.

Moreover, as is the case on the Sun, different parts of a single corona may exhibit different abundances \citep{Sanz-Forcada04, Nordon13}. \citet{Wood06} found that even though two stars may
have very similar properties and activity levels (and being visual binary companions), 70 Oph~A and 70 Oph~B did not exhibit the same coronal abundance patterns.

Given the importance of the heavy elements in the cooling of plasmas with temperatures in the range from 10$^6$ to 10$^{7.5}$ K, it is expected that metallicity could play a significant role in determining the thermal structure (the emission measure distribution, or EMD) of the coronae of such stars. For example, one might expect that in the absence of fractionization mechanisms, coronae with enhanced (depleted) heavy elements would be significantly cooler (hotter) than coronae with solar abundances. 
It is thus of considerable interest to study the coronal spectra of stars whose photospheric abundances differ the most from solar, so as to study how varying the source plasma abundances affects the properties of the coronal plasma that ultimately must originate from the underlying photosphere. 
In the present analysis we select several stars that are metal enriched with respect to  solar from a mean factor of 1.5 to 2.7. We seek to determine whether coronal abundances behave
in a FIP or IFIP behavior, and whether this behavior is a result of depletion or enrichment in either low-FIP or high-FIP elements.

 The present study aims to test whether the FIP (or IFIP) effect can be accurately determined relative to  stellar abundances.
While the standard assumption of solar photospheric abundances may be valid in many cases, a true FIP trend requires the knowledge of true photospheric abundances, which was stressed by Sanz-Forcada et al. (2004).
Photospheric abundances are difficult to measure and to some extent are model dependent, while coronal abundances are comparatively easily measured by counting X-ray photons in emission lines.
The result is that a surprisingly small number of bright coronal sources also have reliable photospheric abundances. We proposed to observe \lm\ after searching the literature for bright X-ray sources that have reliable photospheric abundances that are markedly different than solar.
High-precision photospheric abundances of stars bright enough in X-ray such that well-exposed high-dispersion X-ray spectra can be obtained are few and far between in the literature.

Since different references for solar abundances are used by different authors, it is important to normalize all measurements to a standard solar reference. Abundances presented in this paper are normalized to the solar reference of \citet{Asplund}.

\section{Sample}

\begin{table*}
\caption{Object details}
\label{OBJS}
\centering
\begin{tabular}{cccccccc}

\hline\hline\\[-2ex]
Object & Spectral Type & Distance & Metallicity\footnote{} & References\footnote{} & Total Grating Counts\footnote{} &                            $L_x$                         \\
            &                       &         pc    &   $A_z/A_\odot$          &                                       &                                         & $ 10^{28}\text{erg}\text{s}^{-1}$  \\  
 
\hline \\[-2ex]
\lm\    & G8IV-V & 11.4 & 2.06 &  1,2,3,4,5 & 6588  & 3.2   \\
\iot\   & G0V   & 17   & 1.47 & 6,7,8     & 7785  & 1.6  \\       
HR 7291 & F8V    & 27   & 1.49 & 6,7       & 1345  & 3.7 \\  
\tb     & F7V    & 15.6 & 2.37 &  6,9        & 9741  & 10   \\
\ac\ A  & K1V    & 1.34 & 2.67 & 10,11        & 11840 & 0.1  \\
\ac\ B  & G0V    & 1.34 & 2.15 & 10,11        & 11360 & 0.1   \\
\hline

\multicolumn{7}{l}{\footnotesize 1 Average abundance in photosphere (averaged over all references), with respect to solar \citep{Asplund}}\\
\multicolumn{7}{l}{\footnotesize 2 1:\citet{Soubiran05}, 2:\citet{Prugniel11},
3:\citet{Ramirez07}, 4:\citet{Luck06}, 5:\citet{Milone11}, 6:\citet{Gonzalez07},}\\
\multicolumn{7}{l}{\footnotesize \,\,\,\,7:\citet{Biazzo12},
8:\citet{Bond06}, 9:\citet{Takeda01}, 10:\citet{APrieto04}, 11:\citet{Porto08}}\\
\multicolumn{7}{l}{\footnotesize 3 Zeroth order for \chandra\ }\\

\end{tabular}
\end{table*} 

\subsection{Object selection}
\label{REFS}

Objects for this study were chosen according to two criteria - high photospheric metallicity, and high-statistics X-ray spectra. Sources were filtered from \citet{Gonzalez07} and \citet{APrieto04}. First, objects with supersolar abundances of C, O, Mg, Si, Fe,
and Ni were selected if at least four of these sustained $A_Z/H>1.1$. From these, all objects with X-ray grating observations of at least 1000 photon counts were selected,
leaving five stars, two of which are in a binary system. In addition, \lm\  was observed specifically  in preparation for this analysis \citep[see][]{Drake10}. A summary of objects, their references, and some of their properties is given in Table \ref{OBJS}.

The sample includes six nearby ( <27 pc ) stars with supersolar metallicities ranging from approximately 1.5 to 2.5 solar.
Spectral types of the sample are between K1 and G8. X-ray luminosities span two orders of magnitude, from $10^{27}$ to $10^{29}$ erg s$^{-1}$.
Although different photospheric abundance measurements for these object exist in the literature, they are predominantly  supersolar. 

The  coronal abundances of 11 Leonis Minoris (\lm ) , \iot ologii (\iot ) , and HR 7291 are presented here for the first time, while measurements for the other three, \tb tis (\tb\ )  as well as \ac tauri (\ac ) A and B, are already available in the literature.
The coronal abundances of \tb\ and \ac\ A, B were extensively analyzed by \citet{Maggio11} and by \citet{Raassen03} respectively.  \citet{Maggio11} found a flat FIP dependence, while \citet{Raassen03} measured a solar-like FIP behavior.

\subsection{Additional notes on objects}
\subsubsection{\lm}

A nearby moderately X-ray active star, \lm\ has been erroneously classified as an RS CVn binary by Simbad - probably on the basis of its detection as a rotationally modulated (18.0 days) variable star \citep{Skiff86}. In fact, \lm\ is not a spectroscopic binary according to \citet{Duquennoy91}, with an X-ray luminosity much lower than that of the RS CVn class of close binaries that usually exhibit $L_X = 10^{30 - 32}$ erg s$^{-1}$. It is classified in the spectral class of G8+V in line with an G8 IV-V object.
There is a much fainter ($\Delta V$ = 7.7 mag) visual binary companion 5\arcsec\ away from the G star, probably a mid-M dwarf with a binary orbit of $201$ years  \citep[see][]{Malkov12}. This companion is not the X-ray source since the ROSAT HRI source lies within 1\arcsec of the position of the G star.
\citet{Soubiran05} considered \lm\ to probably be a (95\%) thin-disk star, consistent with its enhanced rotation rate and activity level compared to the Sun. 

\subsubsection{\iot\ }
\iot\ (HD 17051) is a planet-hosting star that is dim in  X-rays. The orbit of the exoplanet is approximately 311 days \citep{Sanz-Forcada13}.
In addition, it has a 1.6 year magnetic activity cycle \citep{Metcalfe10}.
X-ray observations are detailed in \citet{Sanz-Forcada13} where a \xmm\ campaign was used to observe the activity cycle.

\subsubsection{HR 7291}
HR 7291 is an F8V star with a planet with an orbital period of 3 days.
X-ray observations in this case are detailed in \citet{Scandariato13}, as part of a campaign to search for a planet-induced activity cycle.

\subsubsection{\tb\ }
This X-ray bright star hosts a planet with a 3.3-day period. \citet{Maggio11} analyzed the  FIP behavior of this object. No clear FIP (or IFIP) trend was observed. Instead, a flat dependence relative to both solar and actual photospheric abundances was found. In addition, there is a factor 4 of depletion across all elements observed in the corona relative to the photosphere.

\subsubsection{ \ac\ binary}
This spectroscopic binary is part of the nearest stellar system to the Sun (1.34 pc). Both stars are bright X-ray sources.
\citet{Raassen03} observed a FIP effect for both stars, relative to solar abundances.

\subsection{Observations}
The log of observations for the present sample is listed in Table \ref{CTS}. 
Count rates were averaged and exposure durations were summed. Four of these stars were observed in X-rays by \xmm,\ while \ac\ A and B were observed using \chandra. 

The observations of \lm\ , \tb\ , \ac\ A, and B have sufficiently long exposures that result in high-quality grating spectra \citep[see][for details]{Maggio11,Drake10,Raassen03}, making for reliable abundance measurements.
 On the other hand, \iot\ and HR7291 were both observed intermittently during long observation sequences.  As such, the photon count in each individual observation is low (around 200), and many observations exist \citep{Sanz-Forcada13,Scandariato13}. 
 This makes the high-resolution spectra of the Reflection Grating Spectrometer (RGS) noisy, and it is difficult to determine lines. Thus for these two objects the CCD-based European Photon Imaging Camera (EPIC) is more reliable for an abundance analysis (see Sect. \ref{WORK}).

\section{Coronal abundance measurements}
\label{WORK}

We measured the coronal abundances for three of the targets: \lm\ , \iot, and HR 7291. These targets do not have published detailed X-ray abundances.

\subsection{Methodology}\label{meth}
We used the collisional ionization equilibrium (CIE) code VApec \citep{Smith01} within the standard XSPEC11\footnote{http://heasarc.nasa.gov/xanadu/xspec/} software package. 
Models were selected according to the best-fitting reduced $\chi^2$ goodness test, and when several models were considered, the F-Test was used to determine the statistical robustness of the model selection. 
All fitting results, reduced  $\chi^2$ , F-test, and uncertainty values were computed using XSPEC. 
All spectral plots in the paper show data rebinned to a minimal significance of $3\sigma$ for each data point, but not more than three bins, so as to retain the high resolution in the figures.
However, the spectral fits use the original non-binned higher resolution.

In general, the RGS spectra would be preferred because of the high spectral resolution and the ability to visually inspect emission lines.
However, meaningful results can only be obtained for \lm\ (see Sect. \ref {WORK}). When no constraint could be obtained using the RGS, the abundances were measured using the \pn\ spectra. 
 \pn\ is also advantageous for observing Mg and Si, which are visibly discernible and are generally lacking in the RGS spectra.
We concentrated on the well-exposed \pn\ spectrum since it has more than enough counts, and hence we did not use data from the other EPIC-MOS detectors on \xmm .
For \iot\ and HR 7291 none of the RGS data were found useful (again, see Sect. \ref {WORK}).

Joint fits between RGS and \pn\ were considered but discarded. Since the \pn\ provides such a high photon count, the fit is dominated by the \pn\ alone and the high spectral
resolution of the RGS is lost. 
Since both \iot\ and HR 7291 have multiple low-count observations, RGS data are highly noisy and only Fe/O ratios can be constrained.
Therefore only \pn\ was used for these targets. 
A joint fit of all observations was attempted, but significant differences in flux levels between observations rules out a single
model that fits all data well (reduced $\chi^2>2$). 
Instead, each observation was fit separately and results were averaged (see Sect. \ref{WORK}).

\subsubsection{RGS analysis}
When modeling the RGS spectrum (\lm\ only), we used wavelengths between 7 - 37~\AA\ where the S/N is high enough to visibly detect lines. 
The free parameters of each thermal component are the temperature $T$ and the elemental abundances $A_Z$ of the plasma,  as well as the emission measure ($EM = \int n_en_HdV$). 
Testing for multiple $T$ components, for $1T, 2T, 3T$, and $4T
$ the reduced $\chi^2$ is minimal for the $3T$ model with a value of 1.08.
In addition, we conducted an F-test to check for statistical advantage of adding components. As expected, the model with $3T$ model is statistically advantageous (null hypothesis probability of $<10^{-3}$ when compared to any other model).
We thus deem the $3T$ model to be the most appropriate approximation of what is probably a continuous EMD.
All models with two or more temperatures yield consistent relative abundances  to within the errors, which increases our confidence in the derived values.

Abundances were held constant between temperature components assuming abundance dispersion does not change along the EMD.
Since no hydrogen lines are available in X-ray spectra, it is necessary to freeze the abundance of one of the observed elements to avoid the inherent degeneracy between the overall metal abundances and the EM. 
This limits the measurement to relative abundances of the observed elements. 
We chose to fix the Fe abundance to its solar value \citep[again,][]{Asplund}, as Fe-L lines originate from a wide range of $T$.
Hence, we were able to fit for the abundances of C, N, O, Ne, and Ni while all other elements were set to their solar values. 
Indeed, the attempt to fit for the Mg and Si abundances failed,
which forced us to completely freeze their values to solar as well. 
Nevertheless, the $3T$ collisional ionization model reproduces all featured RGS spectral lines well, providing robust results for our free abundance parameters and their uncertainties.
Moreover, setting the Mg and Si values to arbitrarily high or low values has little to no effect on the fit, and no meaningful constraint on these abundances could be derived from the data. 

\subsubsection{\pn\ }
Since the absolute coronal abundances (relative to H) are key to this study because of the supersolar abundances in the photosphere (Sect.~\ref{intro}) and since the RGS falls short in providing absolute values, we must constrain absolute abundances with the \pn\ camera.
While not as accurate as the RGS in determining line positions, the high sensitivity of the \pn\ is advantageous for measuring the bremsstrahlung continuum, which is mostly due to H.
We thus fit the \pn\ spectrum without fixing any of the observable elements to check whether we could obtain absolute abundances that are not degenerate with the normalization (EM).
Since the effective area curve of \pn\ covers the band of $\sim 1.5 - 30$~\AA\ (0.4 -- 10 keV), which is somewhat different from that of the RGS, the model for fitting its spectrum may have a less detailed thermal structure.
We tested both $2T$ and $3T$ models, as suggested by the RGS fits, (for \iot\ and HR 7291 as well).
The third $T$ component does not improve the fit (reduced $\chi ^2$), nor does it reduce the uncertainties on the model parameter values.
Consequently, we preferred the simpler $2T$ model, and all \pn\ results are given for this model.
The Mg and Si emission lines, which are usually lacking in most RGS spectra, are clearly seen around 9~\AA\ and 6.6~\AA\   in all \pn\ spectra we analyzed. 

\begin{table*}
\caption{Observation log}
\label{CTS}
\centering
\setcounter{footnote}{0}
\begin{tabular}{ccccc}

\hline\hline
Object & Date\footnote{} & Instruments & Telescope & Total Duration \\
       &                 &             &           & ks\\ 
 
\hline \\[-2ex]
\lm\        & 09/5/6-7         & RGS, \pn\  & \xmm\ & $85$ \\
\iot\       & 11/5/16-13/2/3   & RGS, \pn\  & \xmm\ & $77$ \\
HR 7291     & 09/9/21-09/10/27 & RGS, \pn\  & \xmm\ & $24$ \\
\tb\        & 03/6/24          & RGS, \pn\  & \xmm\ & $65$ \\
\ac\ A      & 99/12/25         & LETG-HRC-S & \chandra & $82$ \\
\ac\ B      & 99/12/25         & LETG-HRC-S & \chandra\ & $82$ \\
\hline
\multicolumn{4}{l}{\footnotesize 1 Date range represents general times of observations, not one continuous observation.}
\end{tabular}
\end{table*} 

\subsection{Results}
\subsubsection{\lm\ }
\label{LMRES}
\begin{figure}
\centering
\resizebox{\hsize}{!}{\includegraphics{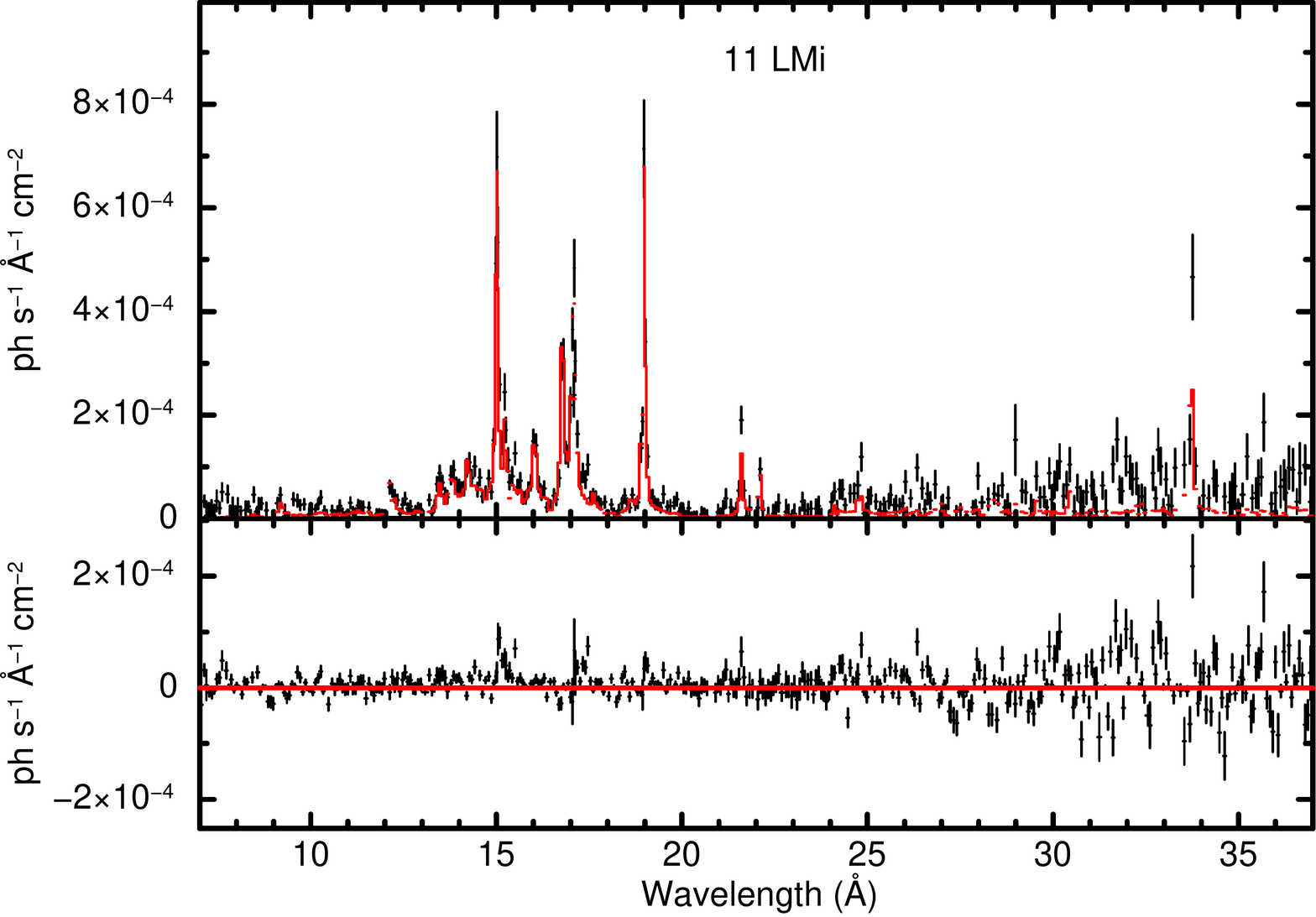}}
\caption{RGS spectrum of \lm\ between 7 -- 37~\AA\ (data points) and best-fit model (red solid line). Residuals are plotted in the bottom panel. The data in the figure are rebinned to $3\sigma,$ but not more than three bins.}
\label{fullspec}
\end{figure}

The full RGS spectrum of \lm\ is presented in Fig.~\ref{fullspec}, along with the residual plot beneath.
An inspection of the spectrum clearly indicates the presence of
the usual main coronal emission lines \citep[e.g.,][]{Behar01}, with the Fe L-shell dominating between 12 -- 18~\AA,  O K-shell at 19 -- 23~\AA , and C K-shell at $\sim 34$~\AA\ being strong
as well.
The Ne-K (12 -- 14~\AA ) and N-K ($\sim 25$~\AA ) lines are weaker. Still, they are easily seen.
At lower wavelengths, Mg-K is just barely detected around 8 -- 9~\AA, while Si-K (6 -- 7~\AA) is not detected at all and is not included in Fig.~\ref{fullspec}.
The best-fit parameters of the model are presented in Table \ref{PNT}. along with their 1$\sigma$ uncertainties. 
The abundance ratios of Fe are fixed to solar in the RGS analysis, but in the \pn\ analysis we find they are consistent with solar values at the 1$\sigma$ level.

The \pn\ spectrum and best-fit model are presented in Fig. 2 and are included in Table \ref{PNT}.
For the most part, all values are  consistent between RGS and \pn\ to within the $1\sigma$ level (except Ni/H at 2$\sigma$), but clearly, some abundance measurements are not very constraining.  
Certainly, the low spectral resolution available with \pn\ and the low S/N available with the RGS could be blamed for this.
We used the RGS absolute abundances of C, N, O, Ne, and Ni, and the \pn\ for Mg, Si, and Fe.

We next wish to ensure that the best-fit solution is not degenerate between its overall metal abundances (metallicity) and the total EM. 
To obtain a confidence contour for the metallicity in the corona of \lm , we tied all best-fit abundance ratios to Fe and also tied the EM ratio of the two components (temperatures were fixed),
effectively leaving only two free parameters - namely the metallicity and EM.
The resulting confidence contours are presented in Fig.~\ref{contour}.
The Fe abundance (with all metals tied to their \pn\ fitted ratio to Fe) is tightly constrained between 0.8 -- 1.1 to $3\sigma$. 
This result gives a tighter constraint, yet is fully consistent with the 0.8 -- 1.5 ($1\sigma$) Fe abundance of the full model, where all abundances were allowed to vary.

\subsubsection{\iot\ and  HR 7291 }
\label{WORK}
For \iot\ and HR 7291 only the \pn\ spectra were used. Temperature variability between observations was the goal and is indeed present during these campaigns. Consequently, a simultaneous fit to all observations with a single 2T CIE model yields an unacceptable fit (reduced $\chi^2>3$). Therefore, each of the 14 (15) observations of \iot\ (HR 7291) were fitted separately. Since we do not find significant variations of the abundances to within $2\sigma$ and the vast majority are consistent to $1\sigma$, we averaged the best-fit abundance values across observations. Admittedly, a few observations are clearly of poor quality, as evident by their reduced $\chi^2>2$ fits or their very large parameter uncertainties. Hence, we cannot rule out abundance variability. Nevertheless, to avoid outliers due to low statistics, the few observations with abundances that are more than $1\sigma$ different from the average were discarded (two for \iot, four for HR 7291).  If the abundances are truly constant, we significantly improve the constraints on the fitted abundances by averaging the results. The standard error was calculated for each elemental abundance and is consistently lower than the statistical errors of each fit (by a factor of 2 at most). Of course, this result is only valid if CIE is prevalent in the plasma for each individual observation.

The results are presented in Table \ref{IOHRGS} for the best-fitting models with average reduced $\chi^2=1.0$ for both \iot\ and HR 7291, including  observations with both high and low reduced $\chi^2$ (>1.2 and <1). Clearly, abundances averaged from all fits provide well-constrained values for all elements, although some could not be resolved by \pn\ or are missing completely, for example, N.
The same method as described in Sect. \ref{LMRES} was used to obtain contour plots between metallicity and EM. We present here a single contour plot for example, since most enclose similar values in parameter space and our final constraints are much better than that of a single observation. Spectra and contour plots are presented in Figs. \ref{iotpn} and \ref{iotcon}. 
The fact that the Fe abundance (representing metallicity) is tightly constrained to $3\sigma$  in all  three analyzed objects to a range of 0.3 - 0.8 in solar units
boosts our confidence in the absolute abundances with respect to H of all elements.

\begin{figure}[!hbt]
\begin{centering}
\resizebox{\hsize}{!}{\includegraphics{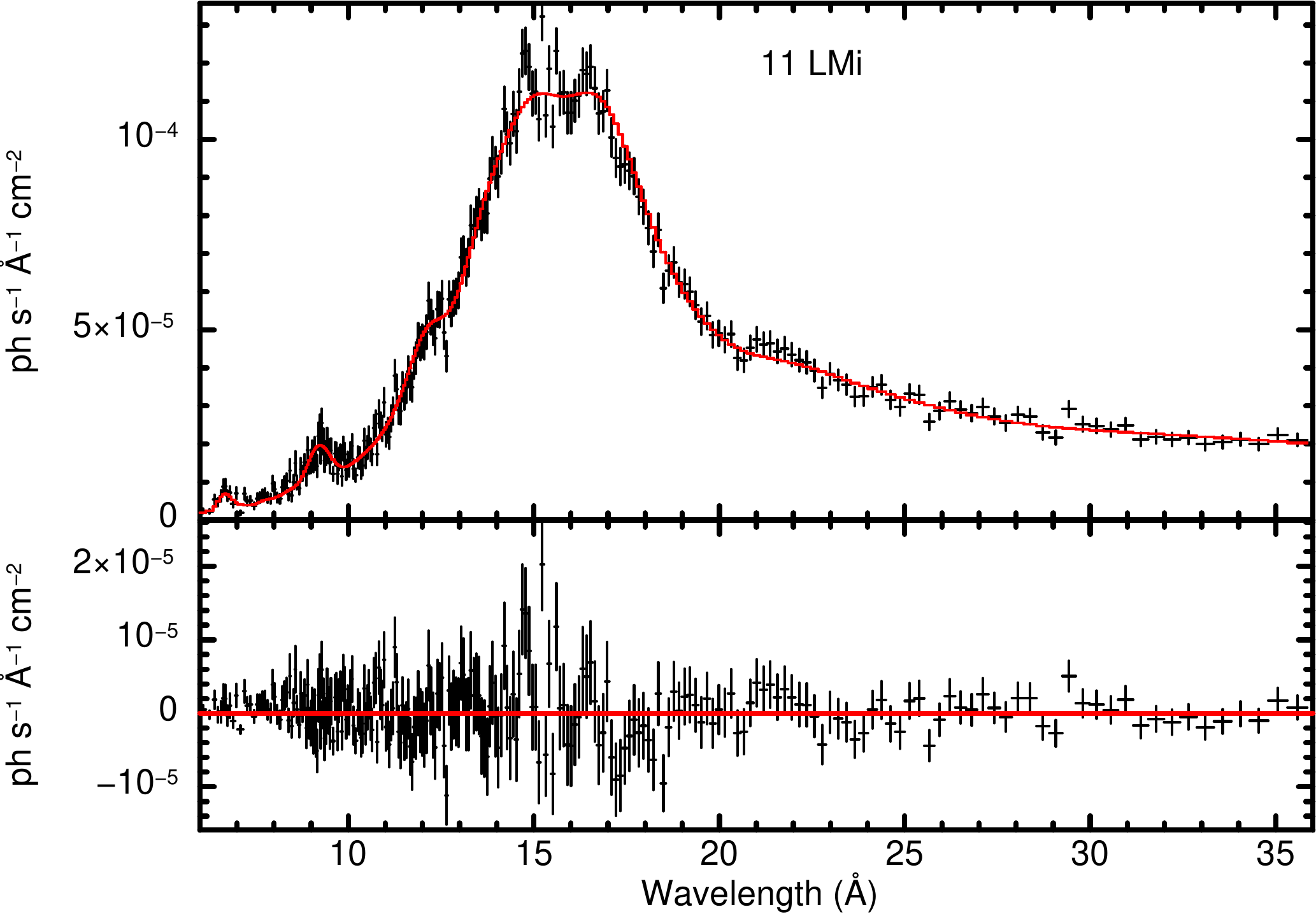}}
\end{centering}
\caption{\pn\ spectrum of \lm\ with best-fit model (in red). Mg and Si emission lines are seen around 9~\AA\ and 6.6~\AA . 
The high statistics allows us to obtain absolute abundances by constraining the line emission (metals) with respect to the continuum (mostly H). Residuals are plotted in the bottom panel. The data in the figures are rebinned to $3\sigma,$ but not more than three bins.}
\label{PNspec}
\end{figure}

\begin{table*}
\caption{\lm\ summary of fitted parameters}
\label{PNT}
\centering
\begin{tabular}{cccccc}
\hline\hline

Parameter & \pn\ Value & Uncertainty & RGS Value & Uncertainty  \\ 
                  &                  & 1$\sigma$   &                  & 1$\sigma$    \\

\hline \\[-2ex]
$\chi^2$       & $273$ & & $5375$ \\
DOF              & $265$ & & $5002$ \\
$kT_1$ [keV] & 0.23 & 0.22-0.24 & 0.14 & 0.11-0.19   \\
$kT_2$ [keV] & 0.57 & 0.55-0.60 & 0.38 & 0.34-0.41  \\
$kT_3$ [keV] &         &                  & 0.69 & 0.57-0.83  \\
&&&&& \\
C [solar] & 0.1 & 0.0-4.2  & 1.1 & 0.6-1.9  \\
N [solar] & 1.5 & 1.0-3.8  & 0.6 & 0.2-1.2  \\
O [solar] & 0.4 & 0.3-0.8  & 0.8 & 0.6-1.0  \\
Ne [solar]& 0.9 & 0.5-2.0 & 0.8 & 0.6-1.0  \\
Mg [solar]& 0.9 & 0.7-1.5 & \nodata & \nodata \\
Si [solar]& 0.6 & 0.4-0.9   & \nodata & \nodata \\
Fe [solar] & 1.0 & 0.8-1.5 & 1.0 &   fixed  \\
Ni [solar] & 4.4 & 2.0-7.1  & 0.8 & 0.0-1.9 \\
\hline

\end{tabular}
\end{table*}

\begin{figure}
\begin{centering}
\resizebox{\hsize}{!}{\includegraphics{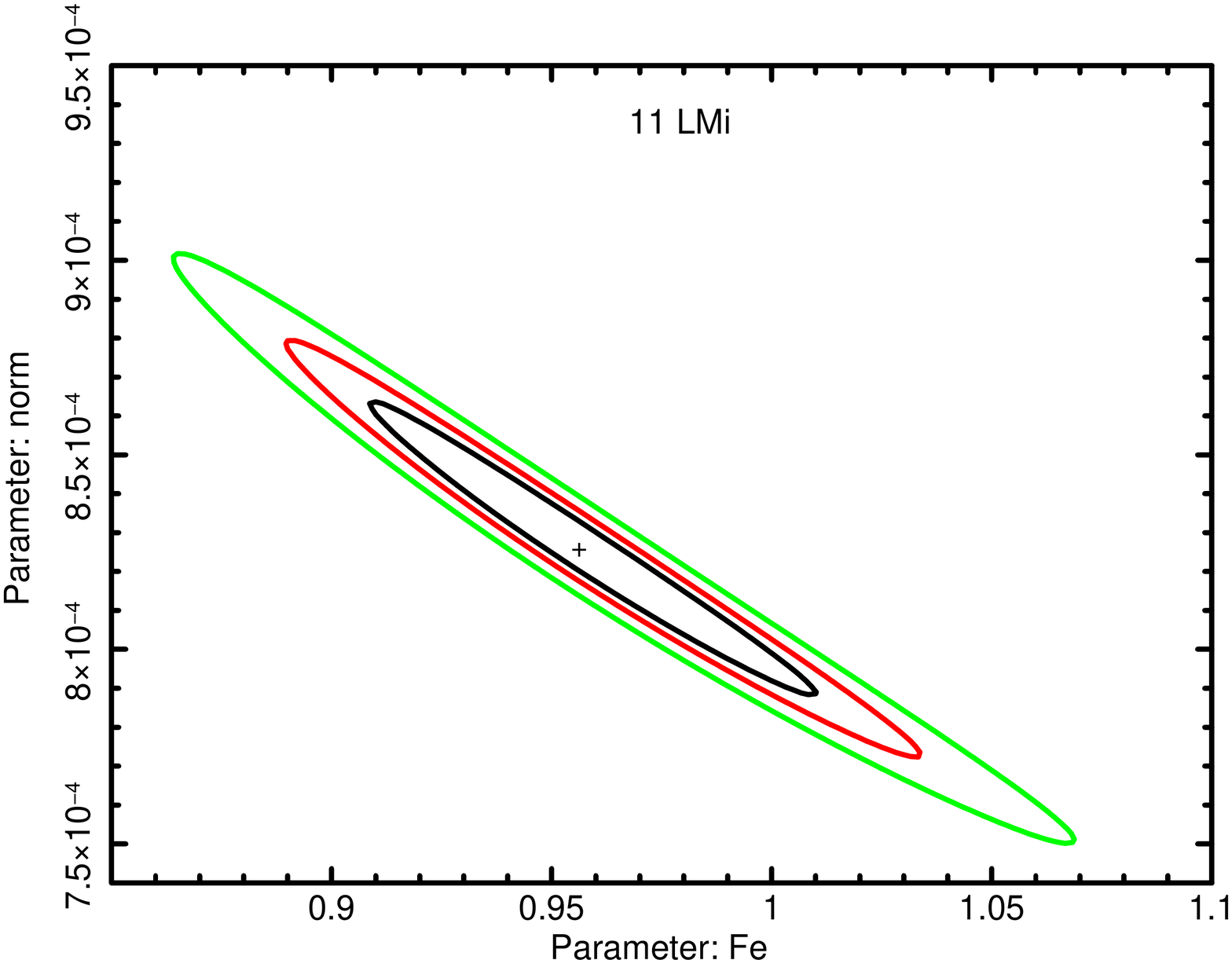}}
\end{centering}
\caption{Contours for \lm\ of overall normalization representing EM (ratio of two thermal components tied) and Fe abundance, while ratios of all other metal abundances to Fe ($A_\text{Z} / A_\text{Fe}$) are set to their \pn\ best-fit values.
The ability to obtain absolute metal abundances with confidence to the level of 3$\sigma$ is demonstrated.}
\label{contour}
\end{figure}

\begin{figure}
    \begin{centering}
    \resizebox{\hsize}{!}{\includegraphics{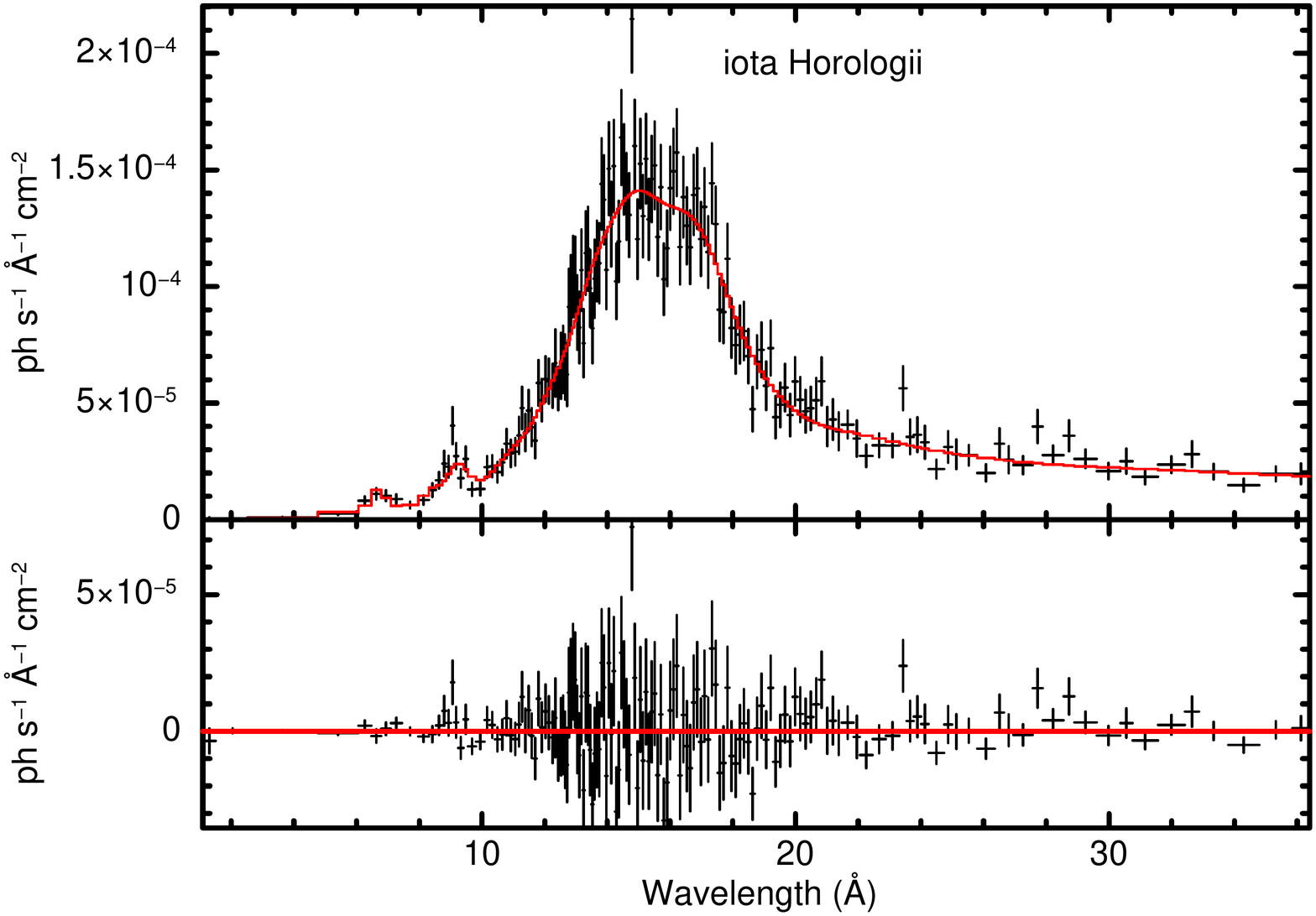}}
    \resizebox{\hsize}{!}{\includegraphics{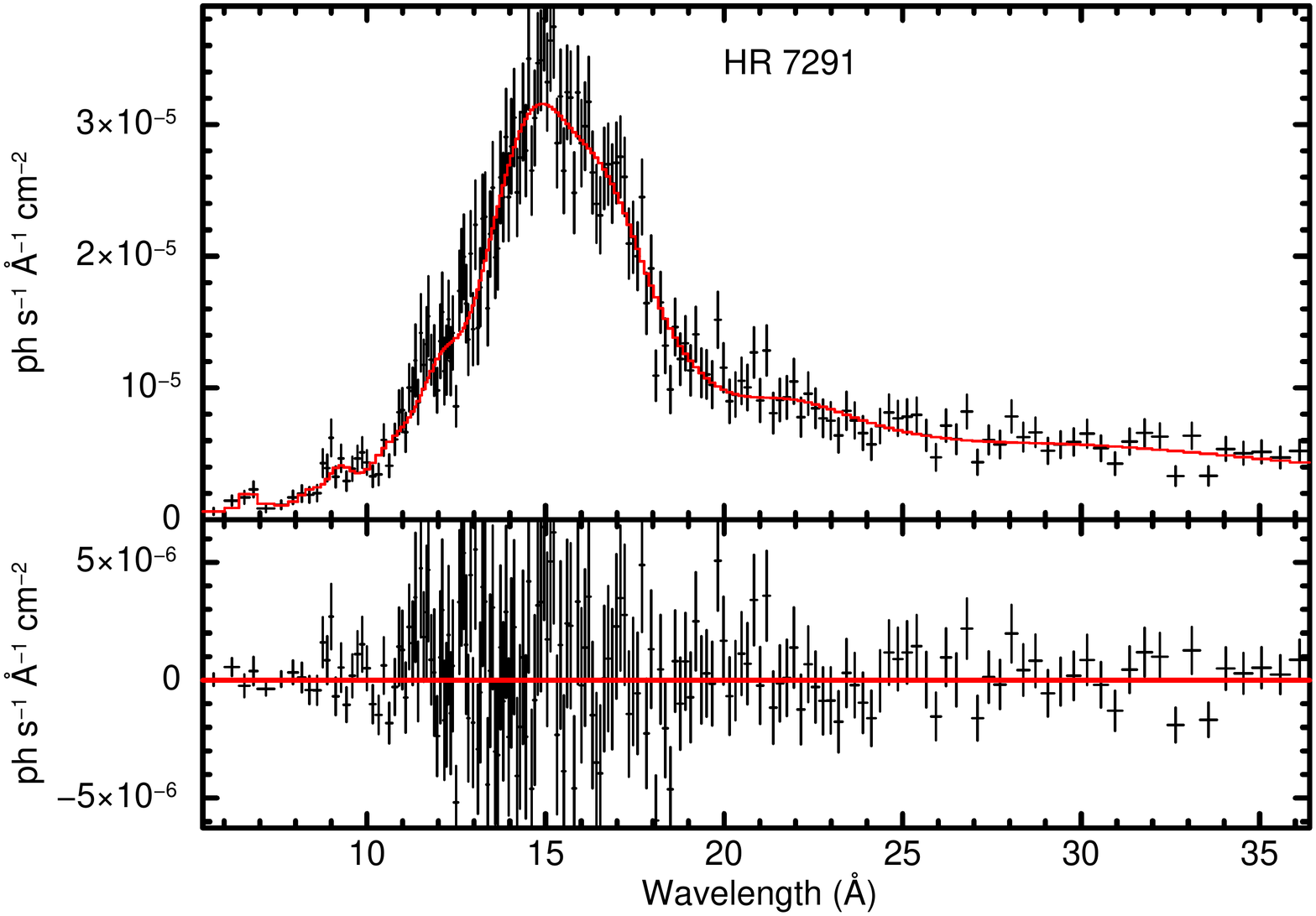}}
\end{centering}
\caption{\iot\ (top) and HR 7921 (bottom) \pn\ spectra, model, and residuals.}
\label{iotpn}
\end{figure}

\begin{figure}
\begin{centering}
    \resizebox{\hsize}{!}{\includegraphics{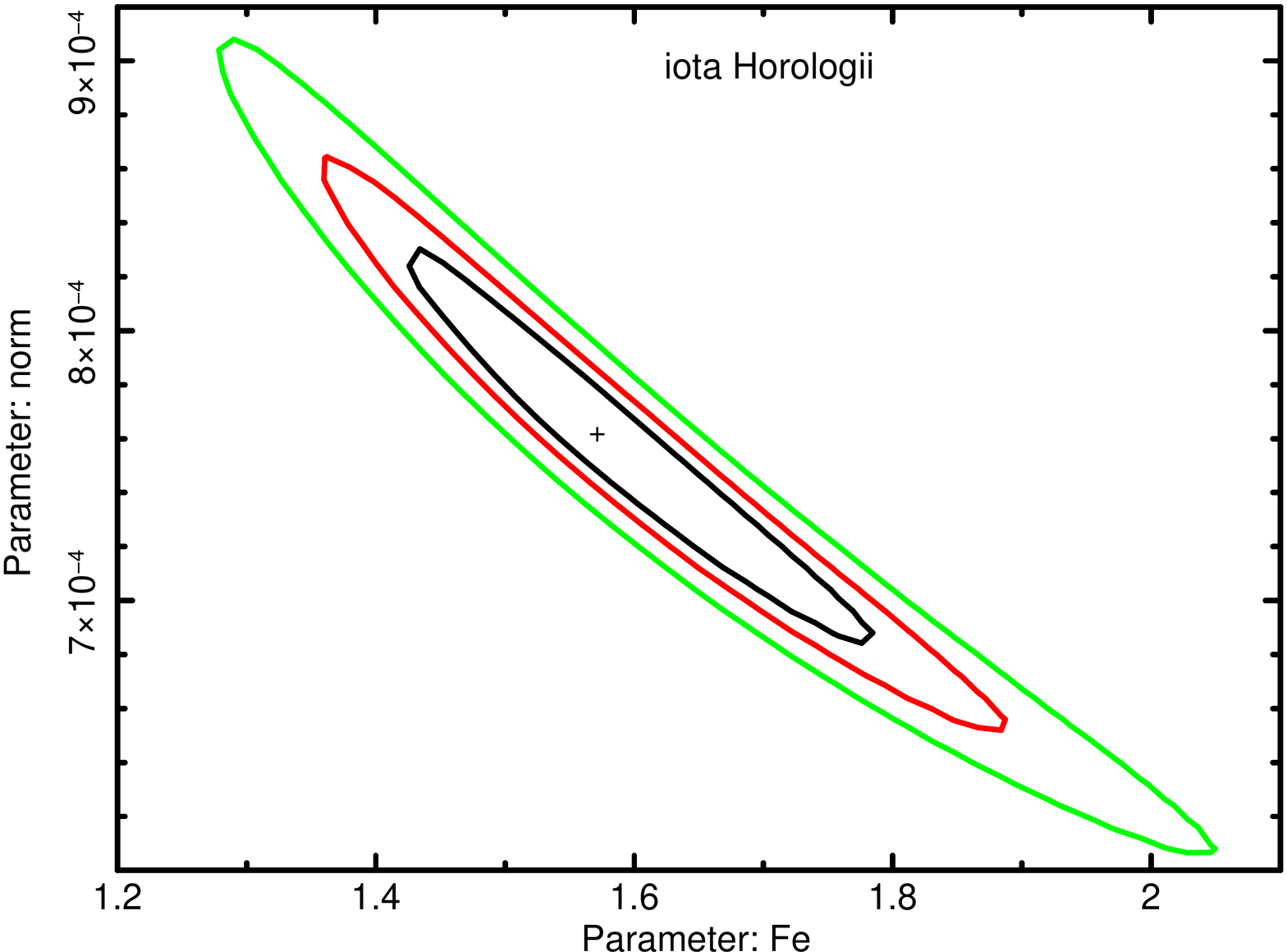}}
    \resizebox{9.65cm}{!}{\includegraphics[angle=90]{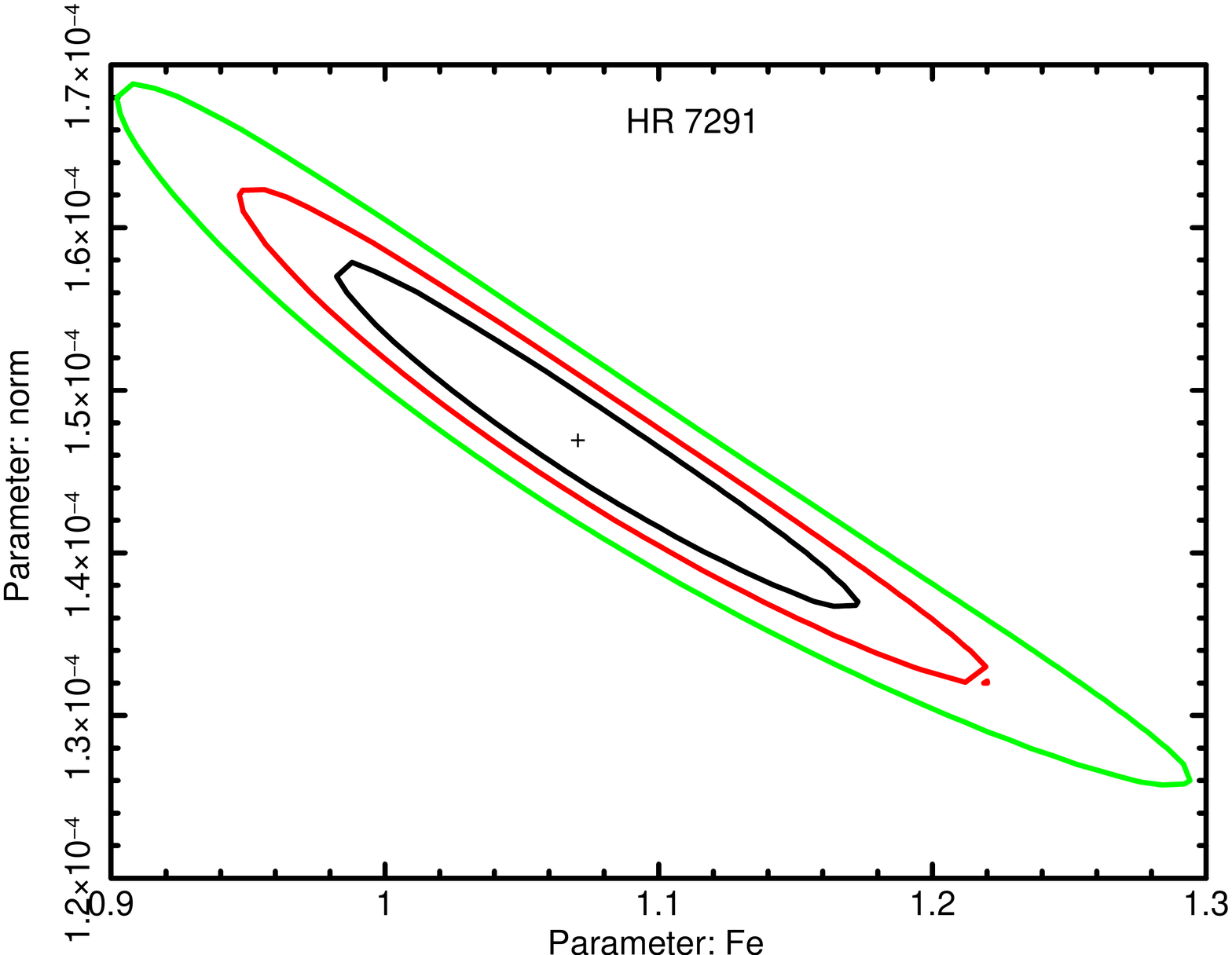}}
\end{centering}
\caption{Example contours of overall normalization representing EM and Fe abundance (metallicity, as in Fig. \ref{contour}) for \iot\ (top) and HR 7291 (bottom). Abundances were constrained much more by averaging across multiple observations.}
\label{iotcon}
\end{figure}

\begin{table*}
\caption{\iot\ and HR 7291 best-fit parameters}
\label{IOHRGS}
\centering
\begin{tabular}{ccccc}

\hline\hline
                  & \multicolumn{2}{c}{\iot\ }  &  \multicolumn{2}{c}{HR7291}  \\
Parameter & Value         & Uncertainty & Value         & Uncertainty\\
                  &                  & 1$\sigma$   &                   & 1$\sigma$  \\ 
 
\hline \\[-2ex]
Avg. $\chi^2$         & $122$ & & $117$ \\
Avg. DOF                & $116$ & & $118$ \\
$kT_\text{1}$[keV] & 0.26 & 0.23-0.30 &   & \\
$kT_\text{2}$[keV] & 0.75 & 0.68-0.81 &   &  \\
&&&\\
O  [solar] & 0.4 & 0.3-0.5 & 0.25 & 0.2-0.3\\
Ne [solar] & 1.1 & 0.9-1.3 & 0.4  & 0.3-0.5\\
Mg [solar] & 1.9 & 1.7-2.2 & 0.9  & 0.8-1.0\\
Si [solar] & 1.5 & 1.2-1.7 & 0.3  & 0.2-0.4\\
Fe [solar] & 1.3 & 1.2-1.4 & 0.8  & 0.7-0.9\\
\hline
\end{tabular}
\end{table*}

\section{Coronal abundances from the literature}
Coronal abundances of \tb\ and \ac\ A and B are only quoted here and not measured again, since high-quality measurements are already available. 
In Table \ref{SUMM} we list these and the above measured abundances and compare them with their respective photospheric abundances taken from the literature (see Table \ref{OBJS}).
Since N and Ne are only measured in the coronae and not in the photospheres, an effective comparison can only be made for five elements - C, O, Mg, Si, and Fe.
All abundances are given relative to \citet{Asplund}. Note that photospheric ranges represent those values found in the literature,
whereas coronal ranges are the measured $1\sigma$ uncertainties. Overall, coronal abundances are significantly lower than their supersolar photospheric counterparts.
In the next section the FIP trends of all these stars are discussed.

\section{Discussion}

\subsection{FIP effect }
 The dependence of  abundances on FIP is plotted in Fig.~\ref{totalfip}. Abundances are normalized to solar \citep{Asplund}, and photospheric values are shown for comparison. The FIP behavior may thus be compared to both solar and actual abundances.
While the uncertainties of the coronal abundances in the figure are those measured in the present analysis or referenced in Sect. \ref{REFS}, the uncertainties of the photospheric data only reflect the range of values in the literature. The lack of real errors on the photospheric values somewhat hinders a meaningful comparison, although by combining several literary sources we obtain a sense of the constraint on these abundances.

The first obvious result observed is that depletion of elements differs significantly when comparing to solar abundances instead of using actual photospheric abundances. 
Furthermore, high-FIP elements are strongly depleted in all stars in the sample. Thus, objects displaying a FIP effect show  little to no depletion of low-FIP elements (unlike the solar corona, which displays an increase of low-FIP elements), and objects displaying no FIP effect show a more even depletion (\lm, \tb\ ). 
For \ac,\ a comparison with solar and actual photospheric abundances leads to the same FIP behavior, but to differing conclusions regarding the origin of the effect.

\lm\ shows no FIP dependence with respect to solar abundances. When comparing with the supersolar photospheric abundances, most significantly Si, but also Ni and possibly Mg are depleted, which may give the impression of an inverse FIP (IFIP) effect. However, Fe, which is a low-FIP element, is not significantly depleted, while O, a typical high-FIP element, is depleted, both of which argue against an IFIP effect. 

In \iot\  the low-FIP elements do not significantly deplete with respect to the photosphere, although they appear to be supersolar. 
O in the corona is significantly depleted with respect to the supersolar photospheric value, producing a solar-like FIP effect.

HR 7291 shows  a rather flat depletion across the FIP axis. There seems to be an enhanced depletion with increasing FIP that creates a solar-like FIP trend as in \iot\ . In this case, the 
depletion of Ne reinforces this effect.
\tb\ is similar to \lm\ as all elements are depleted relative to the photosphere with no I/FIP effect evident. Depletion in this case is much stronger than in the actual photospheric abundances. 

The two \ac\ stars exhibit a stronger FIP effect in their coronae when compared relative to their actual photospheric abundances rather than to solar photospheric abundances, and this is the result of the depletion of high-FIP elements 
rather than the enrichment of low-FIP elements. 
When comparing their coronal abundances to solar photospheric abundances, the FIP effect persists, but now would be due to the enrichment of low-FIP elements, while high-FIP elements remain essentially unchanged.

\subsection{FIP bias and dependence on spectral type}
Continuing our discussion in a more quantitative manner, one can define a FIP bias, for instance, in \citet{Nordon08}, 

\begin{equation}
FB=\log \left( \frac{<A_\text{low}>}{<A_\text{high}>} \right)
\label{FB}
,\end{equation}

\noindent  where $<A_\text{low/high}>$ are the average abundances of low- ($<10$\,eV) / high- ($>10$\,eV) FIP elements. 
Hence, positive/negative $FB$ values reflect a FIP/IFIP bias. 
FIP biases are summarized in Table \ref{FIPB}. Errors are calculated assuming Gaussian distribution and symmetrizing the errors (average of errors if not originally symmetric). We provide three measures here - relative to solar values (using all coronal measurements), relative to existing photospheric values (using only elements measured in both corona and photosphere) and a measure used by \citet{Wood10} (with a negative sign) restricting the FB to the ratio of C, N, O, and Ne (or the available subset) over Fe (solar reference).
\lm\ and \tb\ are consistent with $FB=0$ in all measures. On the other hand, \iot\ , HR7291, and \ac\ A and B exhibit significant FIP bias with $FB>0$ consistent with Fig. \ref{totalfip}. 
The FIP bias with respect to photospheric abundances introduces more uncertainties coming from the photospheric uncertainties, while the other two measures generally have smaller uncertainties because no error on the solar abundances is assumed. Photospheric measurements with only one reference suffer from the same shortcoming.

\begin{figure}
\begin{centering}
    \resizebox{\hsize}{!}{\includegraphics{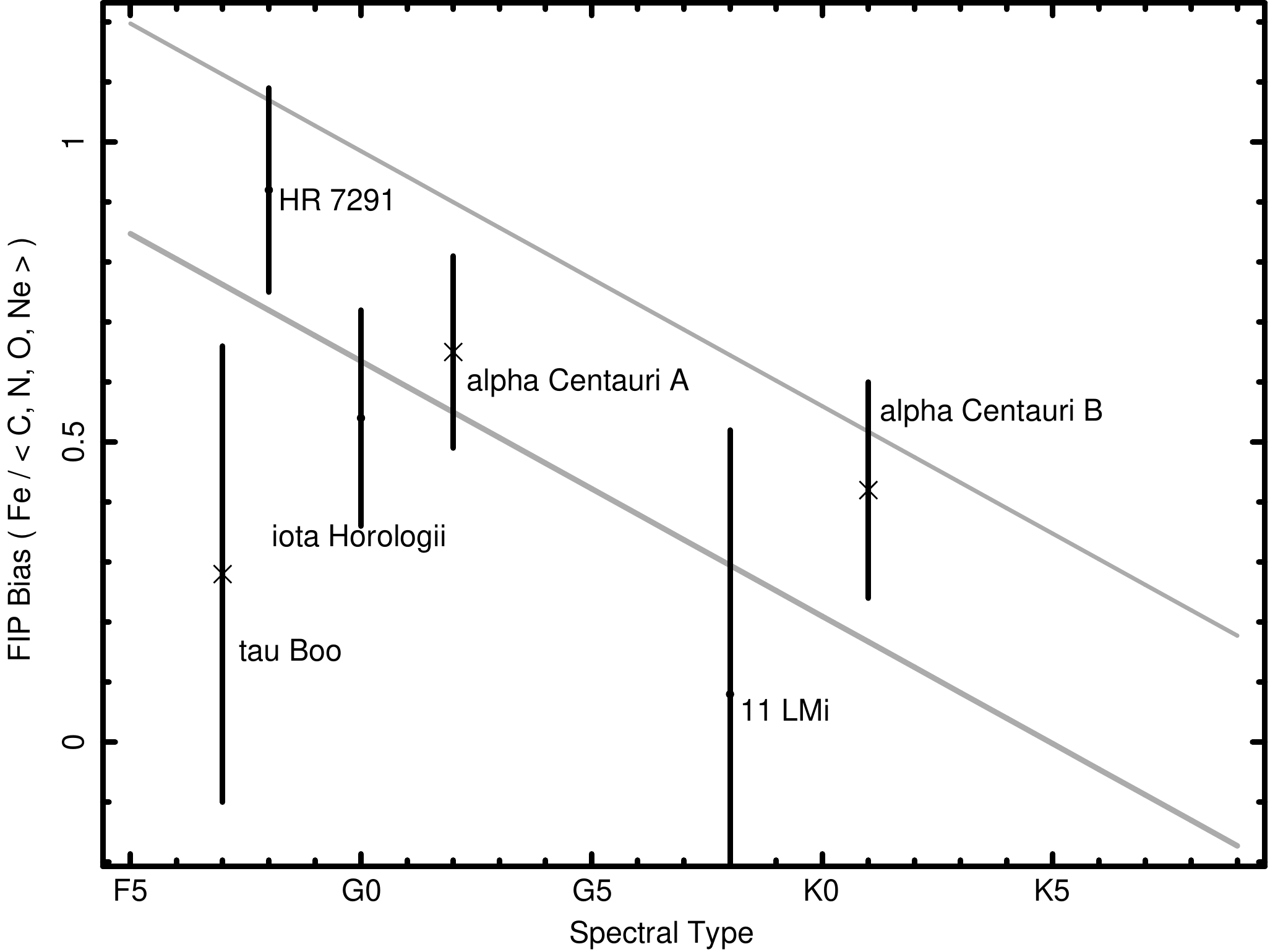}}
\end{centering}
\caption{FIP bias values for the current sample plotted against spectral type, using only C, N, O, Ne, Fe as defined in \citet{Wood10}, only negative. 
Objects marked with an X were already included in \citet{Wood13}. Solid lines bound all of the stars in \citet{Wood12} except for one.  }

\label{FBST}
\end{figure}

The present FIP biases from Col. 3 in Table \ref{FIPB} are compared with the trend reported by \citet{Wood10} in Fig. \ref{FBST}. Note the definition here is the negative of the one defined there. In that paper a  striking correlation of FIP bias with spectral type (Fig. 9 therein) is observed for main-sequence G, K, M dwarfs, which was later extended in \citet{Wood12} and \citet{Wood13}. In the present paper spectral types fall within the range studied in those papers, which have already included \tb\ , and \ac\ A and B.
Though no obvious trend is apparent for the present sample, five of six  (at least marginally) fall within the FIP bias-spectral type correlation band (see Fig. \ref{FBST}).
The outlier is \tb\ (spectral type F6), as already noted in \citet{Wood13} along with Procyon (F5), which lies even farther away ($FB=-0.1$). On the other hand, HR 7291, which is of similar spectral type as Procyon and \tb\  (F8), falls nicely on the correlation (Fig. \ref{FBST}).

\section{Conclusions}

We compared the coronal abundances of six stars with their supersolar photospheric abundances.
For three of them, \lm\ , \iot , and HR7291, we measured absolute coronal abundances from their X-ray spectra. 

\begin{itemize}
        \item{Coronal abundances of all stars are depleted compared to their respective photospheres. Since photospheric abundances are significantly higher than solar, depletion is much greater than when comparing to solar abundance values.}\\
                
        \item{Four of six stars show a positive FIP bias, while the other two have no FIP bias.
        When a FIP effect is present, it is different from the solar FIP effect. 
        In the present sample it appears that the high-FIP elements are depleted, while in the solar corona it is generally accepted that the low-FIP elements are enriched \citep[but this is still disputed, e.g.,][]{schmelz12}.
        }\\

        \item{Five of the stars (all but \tb ) are consistent with a correlation of FIP bias with spectral type \citep{Wood10,Wood13}. 
        }\\
        \item{The importance of knowing the actual photospheric abundances was exemplified by \ac\ A and B. In these two, the FIP bias is the same regardless of photospheric reference. Solar assumptions misleadingly imply enrichment of low-FIP elements, while in fact the high-FIP      elements are depleted. This would lead to different fractionation trends despite the similarity in FIP bias.}
\end{itemize}

In conclusion, high-fidelity measurements of photospheric abundances for additional stars with well-determined coronal abundances would be highly desirable.
Coronal abundance models should  try to explain the phenomena of FIP-independent overall coronal metal depletion in stars with supersolar abundances, as well as a more significant  high FIP element depletion.

\begin{acknowledgements} 
The authors are grateful to an anonymous referee for useful comments that led us to broaden the sample of the paper.
 This research is supported by the I-CORE program of the Planning and Budgeting Committee and the Israel Science Foundation (grant numbers 1937/12 and 1163/10), and by a grant from Israel's Ministry of Science and Technology.
 This work was supported in part by the National Science Foundation under Grant No. PHYS-1066293 and the hospitality of the Aspen Center for Physics.
 \end{acknowledgements}

\begin{sidewaystable*}[h]
\setcounter{footnote}{0}
\caption{Coronal and photospheric abundances$^1$}
\label{SUMM}
\centering
\begin{tabular}{|c|cccccc|cccccccc|}
\hline\hline
            & \multicolumn{6}{c|}{Photospheric Abundances$^2$} & \multicolumn{8}{c|}{Coronal Abundances$^3$} \\
Object & C &O & Mg & Si & Fe & Ni & C   & N & O & Mg & Si & Fe & Ni & Ne \\

\hline \\[-2ex]
\lm\            & \nodata & 2.0     & 0.7-2.4 & 2.6-3.1 & 1.0-1.5 & 2.5-2.8 & 0.6-1.9 & 0.2-1.2 & 0.6-1.0 & 0.7-1.5 & 0.4-0.9 & 0.8-1.5 & 0.0-1.9 & 0.6-1.0 \\
\iot\           & 1.1-2.2 & 2.4     & 1.4-1.6 & 1.2-1.6 & 1.0-2.3 & 1.1-1.5 & \nodata & \nodata & 0.3-0.4 & 1.7-2.1 & 1.2-1.7 & 1.2-1.4 & \nodata & 0.9-1.3 \\
HR 7291         & 1.3-2.4 & 1.0-2.2 & 0.9-1.8 & 1.3-1.6 & 2.5    & \nodata & \nodata & \nodata & 0.2-0.3 & 0.8-1.0 & 0.2-0.4 & 0.7-0.8 & \nodata & 0.3-0.5 \\
\tb\            & 2.9-3.7 & 3.5-2.1 & \nodata & 2.1-2.4 & 1.6-2.0 & 1.7-1.9 & 0.3-0.8 & 0.2-0.7 & 0.2-0.7 & 0.4-0.6 & 0.7-0.9 & 0.4-0.7 & 0.9-3.3 & 0.3-0.5 \\
\ac\ A& 2.6   & 4.1     & 1.7-2.4 & 2.0-3.2 & 1.8-2.8 & 2.2-2.7 & 0.9-1.3 & 0.4-0.9 & 0.4-0.6 & 1.4-1.9 & 1.2-1.7 & 1.4-1.6 & 2.2-3.1 & 0.6-1.2 \\
\ac\ B& 2.0   & 3.1     & 1.9-2.3 & 2.0-2.3 & 1.9-2.0 & 1.4-1.7 & 1.1-1.6 & 0.6-1.1 & 0.5-0.8 & 1.1-1.8 & 1.6-2.3 & 1.3-1.6 & 2.0-2.8 & 0.5-1.2 \\
\hline
\multicolumn{15}{l}{\footnotesize 1 Relative to solar abundance \citep{Asplund}.} \\
\multicolumn{15}{l}{\footnotesize 2 Ranges represent literature disparity: \citet{Soubiran05}, \citet{Prugniel11}, \citet{Ramirez07}, \citet{Luck06}, \citet{Milone11},   \citet{Gonzalez07}.}\\
\multicolumn{15}{l}{\footnotesize \,\,\, \citet{Biazzo12}, \citet{Bond06}, \citet{Takeda01}, \citet{APrieto04}, and \citet{Porto08}. }  \\
\multicolumn{15}{l}{\footnotesize 3 Ranges represent $\pm1\sigma$ intervals from best fit for \iot, \lm, and HR 7291. \tb\ and \ac\ are taken from \citet{Maggio11} and \citet{Raassen03}. } \\

\end{tabular}
\end{sidewaystable*} 

\begin{table*}[h]
\caption{FIP bias measures (uncertainties)}
\label{FIPB}
\centering
\begin{tabular}{|c|ccc|}
\hline
& \multicolumn{3}{c|}{Reference} \\
Object & Photospheric & Solar & \citet{Wood10}\footnote{} \\

\hline \\[-2ex]
\lm\         & 0.08(0.38) & 0.00(0.43) & 0.08(0.44) \\
\iot\        & 0.38(0.20) & 0.31(0.18) & 0.54(0.18) \\
HR 7291 & 0.50(0.41) & 0.32(0.17) & 0.92(0.17) \\
\tb\         & 0.42(0.75) & 0.23(0.54) & 0.28(0.38) \\
\ac\ A     & 0.45(0.22) & 0.37(0.17) & 0.65(0.16) \\
\ac\ B     & 0.34(0.20) & 0.28(0.16) & 0.42(0.18) \\
\hline
\multicolumn{4}{l}{} \\[-2ex]
\multicolumn{4}{l}{ 1 $A_{Fe}/< A_C, A_O, A_N, A_{Ne}>$ }
\end{tabular}

\end{table*} 

\begin{sidewaysfigure*}
\centering
\includegraphics[height=4.1cm]{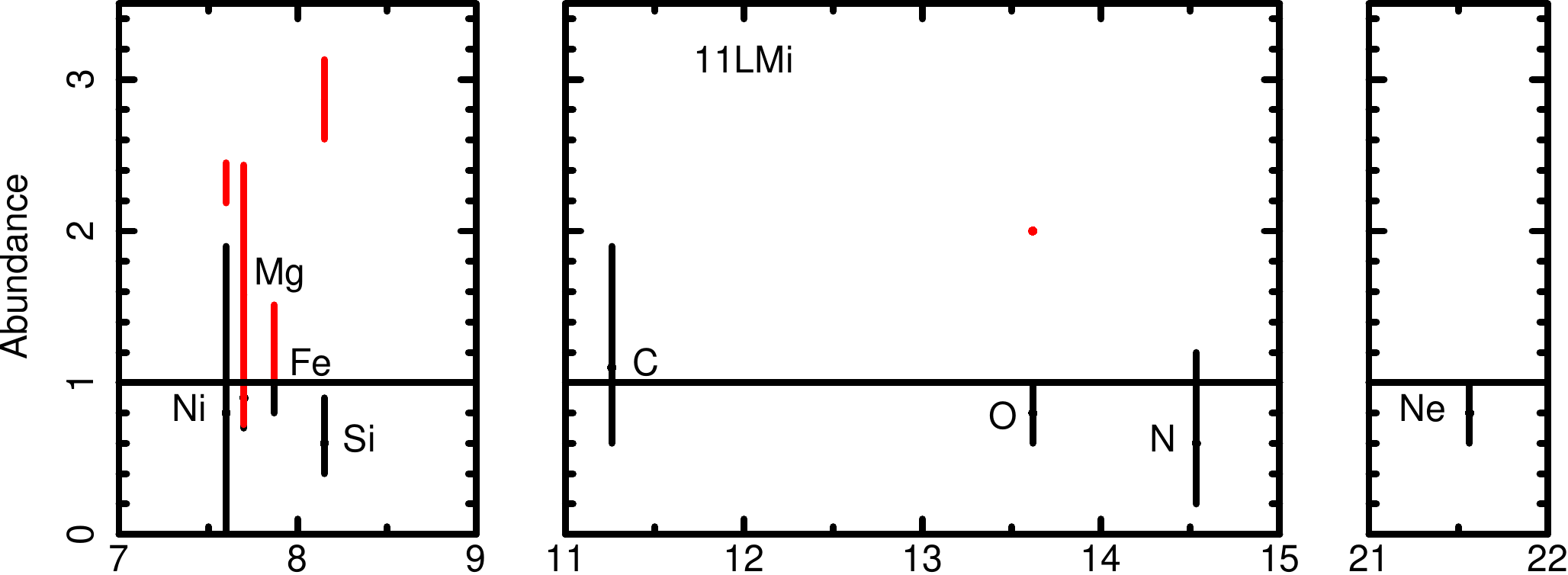} \hspace{2cm}   \includegraphics[height=4.1cm]{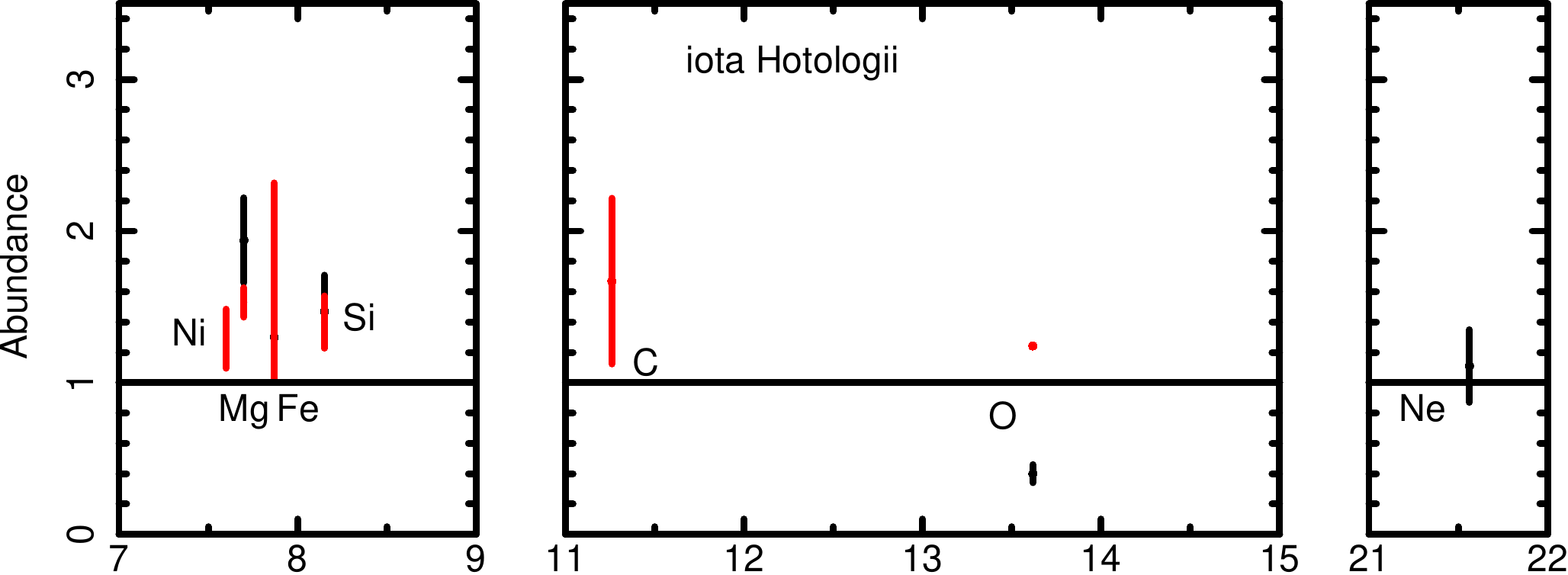}  \includegraphics[height=4.1cm]{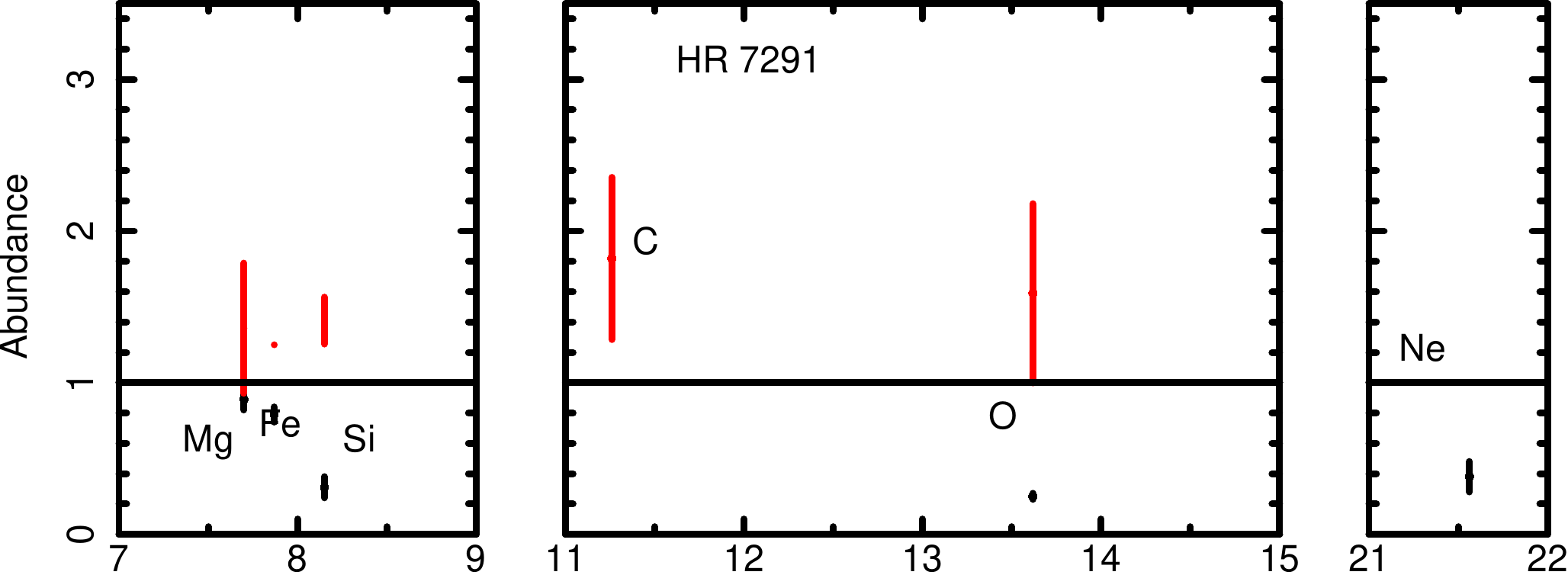}   \hspace{2cm} 
\includegraphics[height=4.1cm]{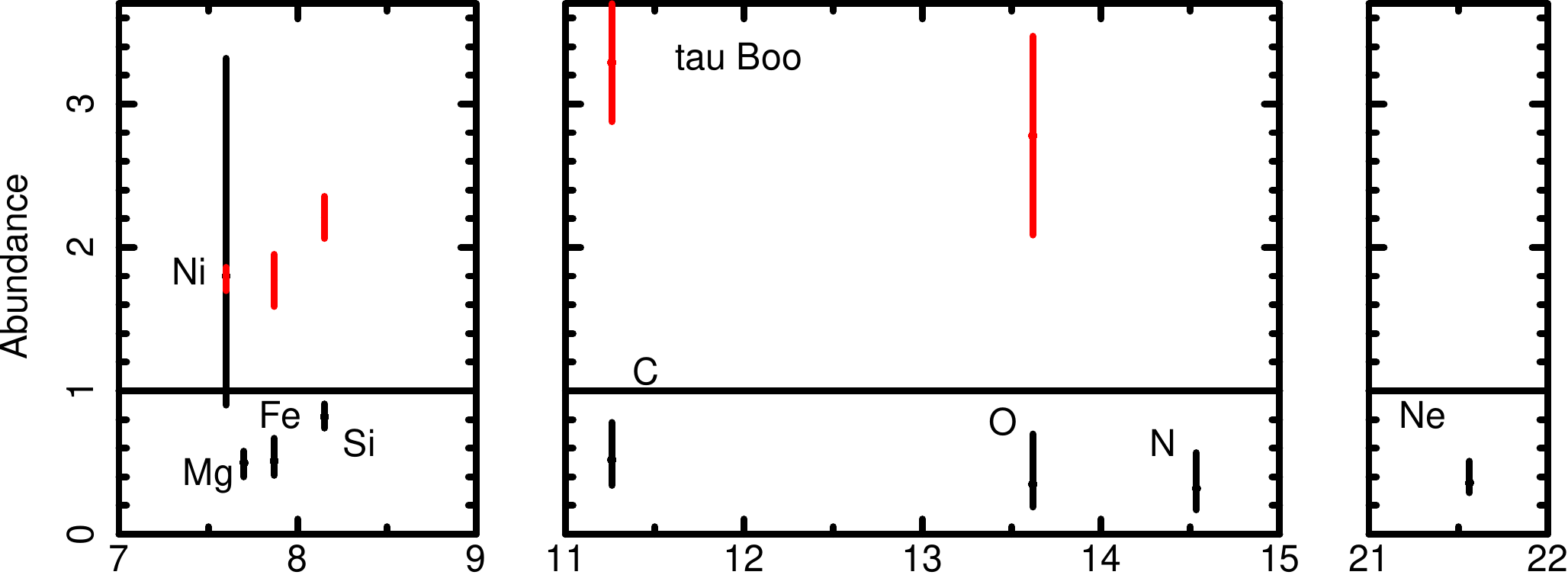} \includegraphics[height=4.1cm]{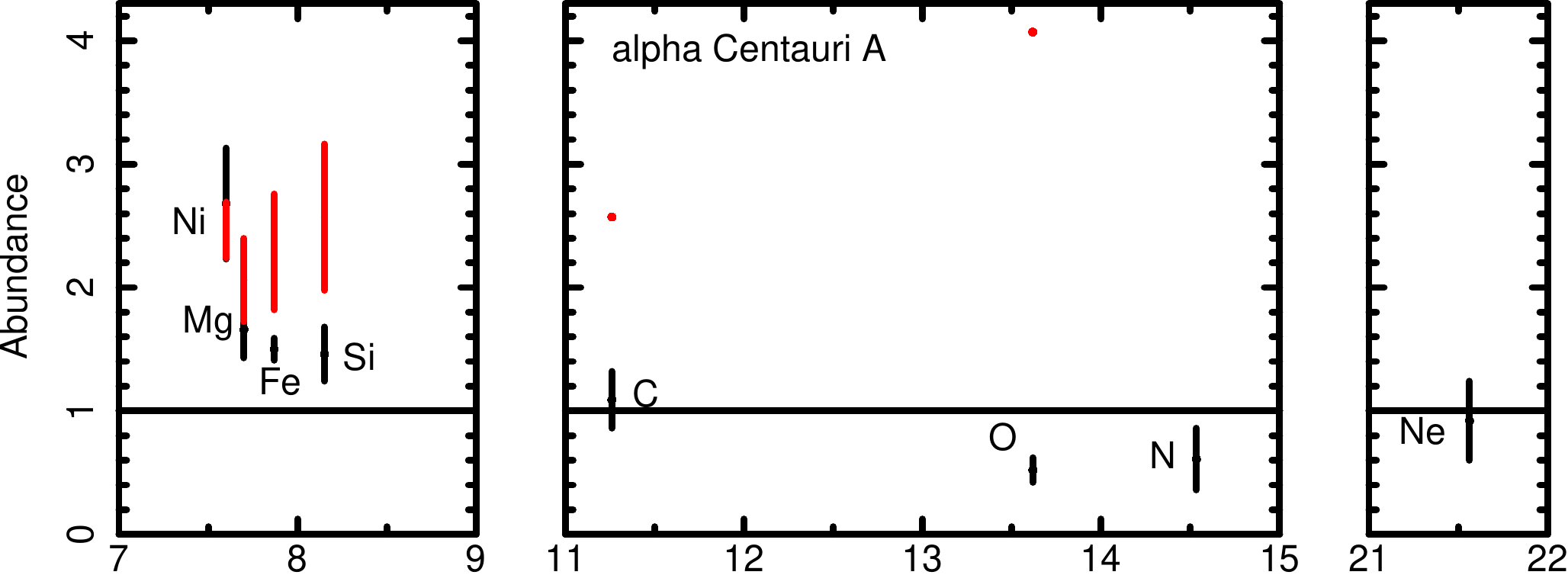}  \hspace{2cm}  \includegraphics[height=4.1cm]{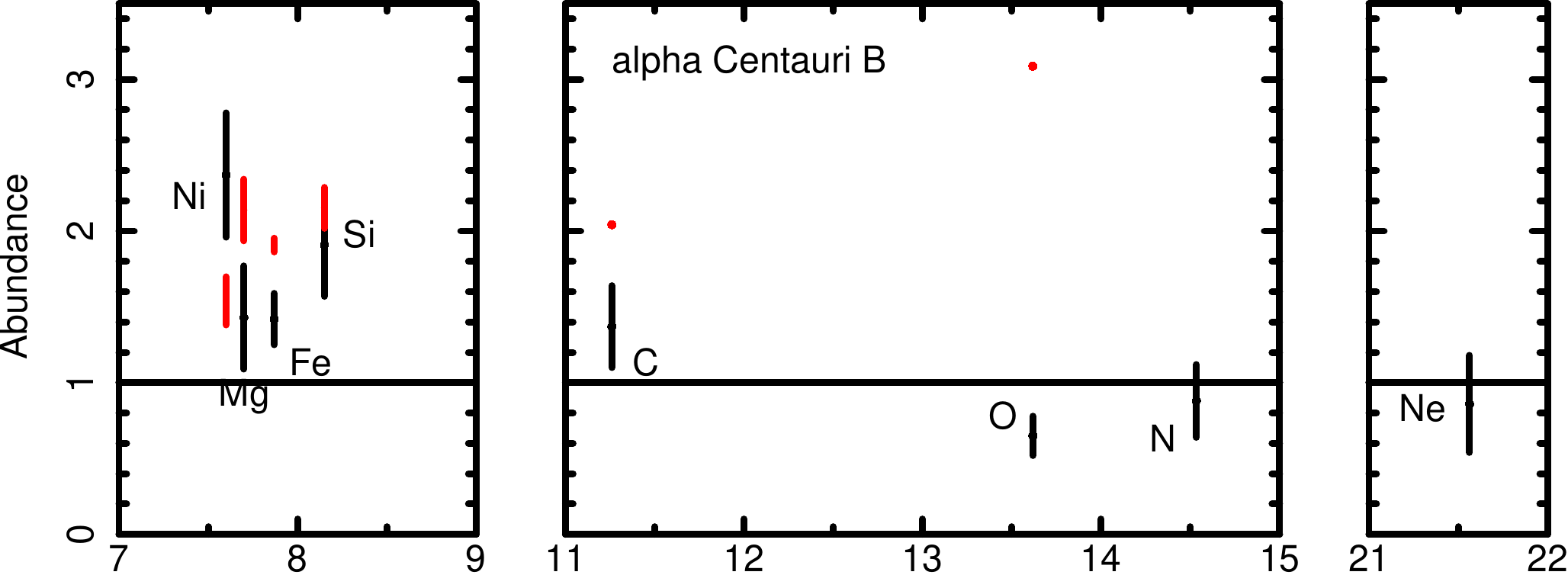}

FIP[eV]
\caption{
Elemental abundances in solar units of the six coronae (black symbols) as a function of the first ionization potential (FIP) compared with the photospheric values (red). Line at 1 represents solar  \citep{Asplund}. Note photospheric abundance ranges represent literature ranges and not actual uncertainties. A FIP effect is observed for \iot , HR 7291, \ac\ A and B, which is not true for \lm\ and \tb .  Clearly, abundance trends relative to solar and relative to actual photospheric values are not always consistent.}
\label{totalfip}
\end{sidewaysfigure*}

\bibliographystyle{aa}

\end{document}